\documentstyle[equations,12pt]{article}

\renewcommand{\theequation}{\arabic{equation}}
\begin{document}
\bibliographystyle{plain}
\newcommand{\ar}{\alpha}
\newcommand{\aar}{\bar{a}}
\newcommand{\bb}{\beta}
\newcommand{\gm}{\gamma}
\newcommand{\Gm}{\Gamma}
\newcommand{\en}{\epsilon}
\newcommand{\dd}{\delta}
\newcommand{\ld}{\lambda}
\newcommand{\oa}{\omega}
\newcommand{\be}{\begin{equation}}
\newcommand{\ee}{\end{equation}}
\newcommand{\bea}{\begin{eqnarray}}
\newcommand{\eea}{\end{eqnarray}}
\newcommand{\bse}{\begin{subequations}}
\newcommand{\ese}{\end{subequations}}
\newcommand{\nn}{\nonumber}
\newcommand{\bR}{\bar{R}}
\newcommand{\bP}{\bar{\Phi}}
\newcommand{\bS}{\bar{S}}
\newcommand{\bU}{\bar{U}}
\newcommand{\bW}{\bar{W}}
\newcommand{\vf}{\varphi}

\title{A time-discretized  version of the Calogero-Moser model}
\author{
Frank W. Nijhoff and \vspace{0.3cm} Gen-Di Pang \\
FB17 Mathematik-Informatik, \\
Universit\"at-Gesamthochschule Paderborn, \\
D-33098 Paderborn, Germany
      }
\date{ }
\maketitle
\begin{abstract}
We introduce an integrable time-discretized  version of the classical
Calogero-Moser model, which goes to the original model in a
continuum limit. This discrete model is obtained from pole
solutions of a discretized version of  the Kadomtsev-Petviashvili
equation, leading to a finite-dimensional symplectic mapping.
Lax pair, symplectic structure and  sufficient set of invariants
of the discrete Calogero-Moser model are constructed. The classical
$r$-matrix is the same as for the continuum model.

\end{abstract}
\vfill \hfill March 1994
\pagebreak

\noindent
{\bf 1. Introduction.}  The original Calogero-Sutherland-Moser model,
\cite{Cal,Suth,Mos}, is a one-dimensional many-body system with
pairwise inverse square interactions, and it is  an integrable model
at both  classical and quantum levels, cf. the two review papers,
\cite{Ols}, for a comprehensive account of both cases. (We will
refer to the classical model as the Calogero-Moser (CM) model).
This long-range interaction model has aroused renewed interest most
recently in view of its close relations to the Haldane-Shastry SU(N)
chain that has been studied in \cite{Hal}. Intriguingly, there is
strong evidence that this integrable model plays an important
role in understanding the universal behaviour in quantum chaos
and mesoscopic physics, \cite{Alt}. Furthermore, rich algebraic
structures, such as a dynamical $r$-matrix, have been found in this
model in \cite{Avan}, cf. also \cite{Skly}.

In this letter we  introduce a time-discretized  version of the
classical CM model. Discrete-time systems are a subject of
intensive research recently, (cf. e.g. \cite{mapp1,mapp2} for
reviews), and they are, in our belief, important in understanding
the true nature of integrability. Most known integrable systems,
such as integrable partial differential equations (Korteweg-de-Vries
equation, Kadomtsev-Petviashvili equation, etc.) or ordinary
differential equations (the Painlev\'e transcendents and elliptic
function equations),  possess, as it has turned out to
be the case, one or more integrable discrete counterparts which
lead to the original model in a well-defined continuum limit.
Finding such integrable discrete versions, however, is in general
a highly non-trivial undertaking. In the present paper, we propose
a new discrete integrable system, namely an exact discrete-time
version of the Calogero-Moser system.

The method used here to obtain the discrete-time  CM model is
based on the observation of \cite{Air}, cf. also \cite{Chu}, that
the dynamics of the poles of special solutions of integrable nonlinear
evolution equations is connected with integrable systems of particles
on the line. The connection between the pole solutions of the
(continuum)  Kadomtsev-Petviashvili (KP) equation and the CM system
was found by Krichever in \cite{Krich}, cf. also \cite{Ves}.  Here we
will perform a similar construction for the discrete case, and we
will show that the pole solutions of a semi-discretized version of
the KP equation  is connected with a time-discretized  version of the
CM model.

\noindent
{\bf 2.  Discrete-time  CM model.} The semi-discretized version of the
KP equation that we will use reads
\begin{equation} \label{eq:1}
(p-q+\hat{u} - \overline{u} )_y=(p-q+\hat{u} -
\overline{u})(u+\hat{\overline{u}}-\hat{u}-\overline{u}),
\end{equation}
where $p$ and $q$ are two (lattice) parameters, $u$ is the (classical)
field, the  ${ \overline{~} \atop {} } $~  denotes the discrete
time-shift corresponding to a translation in the ``time'' direction
while the  $\hat{~}$~ denotes  a shift or translation in the
``spatial'' direction.
In (\ref{eq:1}) the time and spatial variables are discrete while the
third variable $y$ is continuous. Eq. (\ref{eq:1}) is obtained from the
fully-discretized version of the KP equation of \cite{Nij} by letting
one of three lattice parameters tend to zero.

It can be easily proved that  (\ref{eq:1}) is the compatibility
condition of the following two equations
\bse  \label{eq:2}\bea
\overline{\phi} &=& \phi_{y} + (p+ u - \overline{u} ) \phi ,
\label{eq:2a} \\
\hat{\phi} &=& \phi_{y} + (q+ u - \hat{u} ) \phi , \label{eq:2b}
\eea \ese
which is the  Lax representation of (\ref{eq:1}), and the starting
point for the investigation of the integrability characteristics of
the semi-discrete KP.

Direct calculation shows that
\begin{equation} \label{eq:3}
u=\frac{1}{x} \hspace{0.5cm}  {\hbox{with }} \hspace{0.3cm}
x=\frac{n}{p}+\frac{m}{q}+y
\end{equation}
is a (pole) solution to the semi-discretized version of  KP equation
(\ref{eq:1}). Here
$n$ and $m$ are two integers, and $(n , m)$ are considered as the
coordinates of a two-dimensional lattice and
\begin{equation} \label{eq:4}
\overline{x}=\frac{n+1}{p}+\frac{m}{q}+y ,\hspace{0.5cm}
\hat{x}=\frac{n}{p}+\frac{m+1}{q}+y .
\end{equation}
If we substitute this special solution (\ref{eq:3}) into the $u$'s
in (\ref{eq:2}), then we find that
\begin{equation}\label{eq:5}
\phi = (1-\frac{1}{kx})(p+k)^{n}(q+k)^{m} exp(ky)
\end{equation}
satifies (\ref{eq:2}), where $k$ is a  spectral parameter.

Enlightened by the above simple  exercise, we now suppose
{\footnote {We note here that if we take the more restrictive Ansatz
$\phi = \frac{1}{k}( \sum_{i=1}^{N} \frac{b_{i}}{y-x_{i}})  exp(ky) $
instead of that given in (\ref{eq:6}) we still get the same result,
i.e., eqs. (\ref{eq:11}) and (\ref{eq:13}) below.} } that
\bse \label{eq:6} \bea
u &=& \sum_{i=1}^{N}\frac{1}{y-x_{i}} ,   \label{eq:6a} \\
\phi &=& (1-\frac{1}{k} \sum_{i=1}^{N} \frac{b_{i}}{y-x_{i}})(p+k)^{n}
(q+k)^{m}  exp(ky) ,  \label{eq:6b}
\eea \ese
where $x_{i}$ and $b_{i}$ are independent of  $y$ (but they depend on
the time variable),  and we are to find the conditions these $x_{i}$
and $b_{i}$ should satisfy such that equation (\ref{eq:2a}) is valid.
Substituting (\ref{eq:6}) into (\ref{eq:2a}) and
equating to zero the coefficients of $(y-x_{i})^{-1}$ and
$(y-\overline{x_{i}})^{-1}$, we obtain the following equations:
\bse \label{eq:7}\bea
(p+k)b_{i}&=& k+ \sum_{j=1}^{N} \frac{b_{i}}{x_{i}-\overline{x_{j}}}
  -\sum_{j=1 \atop{j \neq i}}^{N}\frac{b_{i}+b_{j}}{x_{i}-{x_{j}}} ,
\label{eq:7a} \\
 (p+k) \overline{b_{i}} &=& k
- \sum_{j=1}^{N} \frac{b_{j}}{\overline{x_{i}}-x_{j}},  \label{eq:7b}
\eea \ese
for all $i=1,2,...,N$,
 where $\overline{b_i}$ and  $\overline{x_i}$  denotes the discrete
time-shift as stated before.
If these  conditions (\ref{eq:7}) are satisfied, then
the $u$ and $\phi$ given by (\ref{eq:6}) satisfy (\ref{eq:2a}).

Now, by introducing the vectors $B=(b_{1}, b_{2},...,b_{N})^{T}$ and
$E=(1,1,1,..., 1) ^{T}$ and the matrices
\bse \label{eq:8}\bea
L_{ij} &=& \left( \sum_{l=1}^{N} \frac{1}{x_{i}-\overline{x_{l}}}
 -\sum_{l=1 \atop{l \neq i}}^{N} \frac{1}{x_{i}-{x_{l}}}\right)
\delta_{ij} - \frac{1}{x_{i}-{x_{j}}} (1-\delta_{ij}) ,
\label{eq:8a} \\
M_{ij} &=& \frac{1}{x_{i}-\overline{x_{i}}} \delta_{ij} -
\frac{1}{\overline{x_{i}}-{x_{j}}} (1-\delta_{ij}) ,  \label{eq:8b}
\eea \ese
we can rewrite  (\ref{eq:7}) in the following form
\bse \label{eq:9}\bea
(p+k) B &=&  k E + L B ,  \label{eq:9a}  \\
(p+k) \overline{B} &=& k E + M B ,  \label{eq:9b}
\eea\ese
and the compatibility of (\ref{eq:9}) leads to the equation
\begin{equation} \label{eq:10}
(\overline{L} M - M L ) B + k ( \overline{L} - M ) E =0 .
\end{equation}
(\ref{eq:10}) is a discrete {\it non-homogeneous}  Lax's  equation.
It can be readily checked that  the resulting
equations of (\ref{eq:10}), i.e.
\begin{equation}\label{eq:11}
 \overline{L} M = M L
\end{equation}
and
\begin{equation} \label{eq:12}
( \overline{L} - M ) E =0  ,
\end{equation}
are consistent and give the {\it same} discrete equations of motion of
a $N-$particle system:
\bea
\frac{1}{x_{i}-\overline{x_i}} + \frac{1}{x_{i}-\underline{x_i}}
+\sum_{j=1 \atop{j \neq i}}^{N}\left( \frac{1}{x_{i}-\overline{x_j}}
                  +  \frac{1}{x_{i}-\underline{x_j}}
   -2  \frac{1}{x_{i}-{x_j}}\right) =0 ,  \nn  \\
{}~~~~~~~~~~~ i=1,2,...,N,  \label{eq:13}
\eea
where $\underline{x_i}$  denotes the discrete time-shift  in the
opposite direction to the one of $ \overline{x_{i}}$.
We will call the model, for which eq. (\ref{eq:13}) are the equations
of motion, the discrete-time CM model. We will
show below that in a continuum limit these equations go to that of the
original  CM model.

The Lax pair for this discrete-time CM model are given by $L$ and $M$
of (\ref{eq:8}).
{}From eq. (\ref{eq:10}), it can be readily seen that
\begin{equation} \label{eq:14}
 \overline{I_{k}} = I_{k}
\end{equation}
for any $k=1,2,...$, where
\begin{equation}\label{eq:15}
I_{k} \equiv Tr ( L^{k}) ,
\end{equation}
leading to a sufficient number of invariants (or conservation laws)
of the discrete-time flow given by (\ref{eq:13}). \\
\\
\\
{\bf 3. Symplectic Structure.} In order to establish the exact
integrability of the discrete CM model (\ref{eq:13}), we need first to
establish an appropriate symplectic structure for the N-particle
system. The discrete-time flow will then have the interpretation of
the iterate of a canonical transformation with respect to that
symplectic structure. To get the generating function of this
canonical transformation, we will follow the point of view of
ref. \cite{RS}. We start by noting that eq. (\ref{eq:13}) can
actually be obtained from the variation of a discrete action,
given by
\begin{equation} \label{eq:16}
S = \sum_n {\cal L}(x,\overline{x}) =
\sum_n \left( \sum_{i,j=1}^{N} \log | x_{i}-\overline{x_{j}}|
 -\sum_{i,j=1 \atop i \ne j }^{N} \log | x_{i}-{x_{j}}| \right)   ,
\end{equation}
in which the sum over $n$ denotes the sum over all discrete-time
iterates. The discrete Euler-Lagrange equations
\be\label{eq:16a}
\overline{\frac{\partial {\cal L}}{\partial x_i}} +
\frac{\partial {\cal L}}{\partial \overline{x_i}} = 0  , \hspace{0.5cm}
i=1,2,..,N,
\ee
yield the equations of motion  (\ref{eq:13}).

The form of the `kinetic' term in our action (\ref{eq:16})
invites to perform a Legendre transformation in the following form
\begin{equation}\label{eq:17}
{\cal H}(\overline{p},x) =\sum_{i,j=1}^{N} \overline{p_{ij}} (
                x_{i}-\overline{x_{j}}) - {\cal L}(x,\overline{x})\  ,
\end{equation}
from which we obtain
\begin{equation} \label{eq:18}
\sum_{i=1}^{N} \overline{p_{ij}}= -\frac{\partial {\cal L}}{\partial
\overline{x_{j}}} ,
\end{equation}
and
\bse \label{eq:19}\bea
\frac{\partial {\cal H}}{\partial x_{i}} &=&  \sum_{j=1}^{N}
(\overline{p_{ij}} -p_{ji}),  \label{eq:19a} \\
\frac{\partial {\cal H}}{\partial \overline{p_{ij} }} &=&
x_{i}-\overline{x_{j}}.  \label{eq:19b}
\eea \ese
Eqs. (\ref{eq:19a}) and (\ref{eq:19b}) can be interpreted as
discrete Hamilton equations, but, of course, ${\cal H}$ is not
a Hamiltonian in the usual sense of the word. ${\cal H}$ is the
generating function of the canonical transformation $x\mapsto
\overline{x}$, $p\mapsto \overline{p}$. This transformation, as
a consequence of eqs. (\ref{eq:19}), will leave the following
symplectic form invariant:
\begin{equation} \label{eq:20}
\Omega = \sum_{i,j=1}^{N} dp_{ij} \wedge dx_{j}  ,
\end{equation}
for which we have $ \overline{\Omega} =\Omega$.

Eqs. (\ref{eq:17})-(\ref{eq:20}) are still general. Coming back
to the special case of the action (\ref{eq:16}) we will find from
(\ref{eq:18}) that
\begin{equation} \label{eq:21}
\overline{p_{ij}}=\frac{1}{x_i-\overline{x_{j}}}
\end{equation}
and from (\ref{eq:17}) we find ${\cal H}$:
\be \label{eq:22}
{\cal H}(\overline{p},x) = \sum_{i,j=1}^N \log | \overline{p}_{ij} |\,+\,
\sum_{i,j=1\atop i\ne j}^N \log | x_i - x_j |\   .
\ee
The symplectic structure (\ref{eq:20}) leads to the following
Poisson brackets
\begin{equation}  \label{eq:23}
\sum_{i=1}^{N} \{p_{ij},x_{k}\}=\delta_{jk}  .
\end{equation}

\noindent
{\bf 4. Classical r-matrix.} In order to prove the complete
integrability of the discrete-time CM model, we now construct its
$r$-matrix structure. From the above discussion it follows that
we can make the following choice of canonical variables $(p_i, x_i)$
of the $i$th particle
\be\label{eq:24}
p_i = \sum_{j=1}^N p_{ji} + \sum_{j=1\atop j\ne i}^N \frac{1}{x_i - x_j}
\  ,
\ee
leading to the standard Poisson brackets
\begin{equation} \label{eq:25}
\{p_i, x_j\}=\delta_{ij}, \hspace{0.4cm}  \{p_i, p_j\}=\{x_i, x_j\}=0.
\end{equation}
In terms of these canonical variables, the Lax matrix
$L$ can be written as:
\begin{equation} \label{eq:26}
L=\sum_{i=1}^{N} p_i e_{ii} - \sum_{i,j=1 \atop i\ne j}^{N}
\frac{e_{ij}}{x_i -x_j} ,
\end{equation}
where the elements of the matrix $e_{ij}$ are defined as
$(e_{ij})_{kl}=\delta_{ik} \delta_{jl}$. Eq. (\ref{eq:26})
is the usual $L$-matrix for the CM model, and thus we can immediately
use the result of \cite{Avan} in order to establish the involutivity
of the invariants (\ref{eq:15}). In fact, using (\ref{eq:25}) and the
expression for $L$, we can calculate the fundamental Poisson bracket
structure in terms of the matrices $L$, leading to the well-known
result
\begin{equation} \label{eq:27}
\{L\stackrel{\otimes}{,}L\}= [ r_{12}, L \otimes 1 ] - [ r_{21}, 1  \otimes L
],
\end{equation}
where $\otimes$ is the tensor product, $[\cdot, \cdot]$ denotes
the usual commutator and
$$
\{L\stackrel{\otimes}{,}L\} \equiv \sum_{i,j=1 }^{N} \sum_{k,l=1 }^{N}
\{ L_{ij}, L_{kl}\} e_{ij} \otimes e_{kl} .
$$
The $r$-matrix in (\ref{eq:27}) is given by, \cite{Avan},
\begin{equation} \label{eq:28}
r_{12} = \sum_{i,j=1 \atop i\ne j}^{N} \frac{1}{x_j -x_i} e_{ij} \otimes e_{ji}
+\frac{1}{2} \sum_{i,j=1 \atop i\ne j}^{N} \frac{1}{x_j -x_i} e_{ii}
\otimes (e_{ij}-e_{ji}) ,
\end{equation}
obeying $r_{21}= P r_{12} P$, where $P$ is the permutation matrix:
$P x \otimes y P = y \otimes x $.

Thus, in going from the continuous to the discrete CM model, both
the $L$-operator as well as the $r$-matrix remain the same. What changes
is the $M$-matrix, which will now depend on the original momentum
variables $p_{ij}$ of eq. (\ref{eq:21}).

The involutivity of the invariants
\begin{equation} \label{eq:29}
\{ I_k, I_l \}= \{ Tr (L^{k}), Tr (L^{l}) \} =0 \hspace{0.7cm}
{\hbox{for all}}\hspace{0.3cm} k,l=1,2,....\  ,
\end{equation}
that follows as a consequence of the $r$-matrix structure (\ref{eq:27}),
will lead to the integrability of the discrete-time model by an
argument presented elegantly in \cite{RS}, forming the discrete
counterpart of the Arnol'd-Liouville theorem.
Therefore, we have proved the complete integrability of
the discrete-time CM model.\\
\\
\\
{\bf 5. Continuum Limit.}  We now show that in a continuum limit
(\ref{eq:13}) goes to that of the original CM model. In fact, the
considerations above show that the invariants and the canonical
variables remain the same on the continuous- as well as the discrete-
time level. Thus, taking the invariants (\ref{eq:15}) as Hamiltonians,
we get a hierarchy of continuous flows interpolating the
discrete-time flow, the one of order $k=2$ corresponding to the
original model. In order to perform the continuum limit, we first set
\begin{equation}\label{eq:30}
  x_{i}= z_{i} + n \alpha , \hspace{0.4cm} i=1,2,..., N,
\end{equation}
where $\alpha$ is a small (constant) parameter and
\begin{equation}\label{eq:31}
  \overline{x_{i}}= \overline{z_{i}} + (n+1) \alpha , \hspace{0.4cm} i=1,2,...,
N.
\end{equation}
 Then, in the continuum limit, we write
\begin{equation}\label{eq:32}
  \overline{z_{i}}= z_{i} + \epsilon \dot{z_{i}} + 1/2 \epsilon^{2}
 {\ddot z_{i}} +...   ,
\end{equation}
for all $i=1,2,..,N$, where $\dot{z_{i}} \equiv \frac{d}{dt} z_{i} $ and
$\epsilon $ is the time-step parameter which, we suppose, is in the
order of $ O(\alpha^{2})$.    Substituting (\ref{eq:30})-
(\ref{eq:32}) into (\ref{eq:13}), we get as the leading
order term of (\ref{eq:13}), i.e.
\begin{equation} \label{eq:33}
 {\ddot z_{i}} =-2g \sum_{j=1 \atop j \ne i}^{N} \frac{1}{(z_{i}-z_{j})^{3}},
 \hspace{0.4cm} i=1,2,..., N,
\end{equation}
where $g \equiv \ar^4/ \epsilon^{2}$. It is clear that
(\ref{eq:33}) are exactly the equations of motion of the
(continuous) CM model. It is interesting to note that the
coupling constant of the continuous model arises from the
discrete-time step. \\
\\
\\
{\bf 6. Disscusion.}
In this letter, we have constructed a discrete-time CM model. This
discrete model is also integrable and is the iterate of a canonical
transformation generated by the discrete `Hamiltonian'
 (\ref{eq:22}).
The original CM model is a limit of this discrete-time CM model, and
in this limit the coupling constant for the long-range interaction
term is encoded in the discrete-time step parameter.
The present result invites a number of
interesting problems to be studied.
First of all, one should address the generalization of these
results to the more generic elliptic potentials, cf. \cite{Krich,Skly}.
Secondly, one may ask whether one can find time-discretizations of
the relativistic version of the CM model, \cite{Ruijss}, cf. also
\cite{BC}. In fact,
Suris, in \cite{Suris}, has found an interesting connection between
the discrete-time Toda model and its relativistic version. Furthermore,
an intriguing connection exists between the sine-Gordon soliton
solutions and the relativistic CM model, cf. \cite{Bab}. This could
lead to an alternative way to discretize the model. Thirdly, for the
discrete-time model, following the similarities with the structure of
the integrable quantum mappings studied in \cite{NCP}, one should
investigate not only the $L$-part of the Lax pair, but also take the
$M$-part under consideration in the $r$-matrix structure. This has
been done for the integration of mappings of KdV type in
\cite{finegap}.
Finally, these investigations should also be pursued on the quantum
level. Although, the discrete-time model has an obvious quantum
counterpart, in much the same way as the continuum model,
the work of \cite{NCP} on quantum mappings, as well as the
work \cite{Wadati} on the quantum CM model, indicate that some
important modification (e.g. with respect to the construction of
exact quantum invariants of the discrete-time flow) might be
expected. All these problems are being investigated and will be
discussed in detail in future publications.
\vspace{.5cm}

\noindent
\subsection*{Acknowledgement}

Both of us would like to thank  Prof. B. Fuchssteiner for his warm
hospitality and  the Alexander von Humboldt Foundation for financial
support. FWN thanks Jorge Jose for bringing ref. \cite{Alt} to his
attention.

\pagebreak

\end{document}